\documentclass[pra,superscriptaddress,floatfix,twocolumn,longbibliography]{revtex4-1}
\usepackage{graphicx}
\usepackage{wasysym}
\usepackage{color}

\def\lsco{La$_{2-x}$Sr$_x$CuO$_4$}
\def\lbco{La$_{2-x}$Ba$_x$CuO$_4$}

\def\lcod{La$_2$CuO$_{4+\delta}$}

\begin{document}

\title{Neutron scattering study of spin ordering and stripe pinning in superconducting La$_{1.93}$Sr$_{0.07}$CuO$_4$}

\author{H. Jacobsen}
\affiliation{Condensed Matter Physics \&\ Materials Science Department, Brookhaven National Laboratory, Upton, NY 11973-5000, USA}
\affiliation{Nano-Science Center, Niels Bohr Institute, University of Copenhagen, DK-2100 Copenhagen \O, Denmark}
\author{I. A. Zaliznyak}
\affiliation{Condensed Matter Physics \&\ Materials Science Department, Brookhaven National Laboratory, Upton, NY 11973-5000, USA}
\author{A. T. Savici}
\affiliation{Neutron Data Analysis and Visualization Division, Oak Ridge National Laboratory, Oak Ridge, TN 37831, USA}
\author{B. L. Winn}
\affiliation{Quantum Condensed Matter Division, Oak Ridge National Laboratory, Oak Ridge, TN 37831, USA}
\author{S. Chang}
\thanks{Present address: 
Physics Today, One Physics Ellipse, College Park, MD 20740-3842, USA.}
\affiliation{NIST Center for Neutron Research, National Institute of Standards and Technology, Gaithersburg, MD 20899, USA}
\author{M. H\"ucker}
\author{G. D. Gu}
\author{J. M. Tranquada}
\email{jtran@bnl.gov}
\affiliation{Condensed Matter Physics \&\ Materials Science Department, Brookhaven National Laboratory, Upton, NY 11973-5000, USA}
\date{\today}
\begin{abstract}
The relationships among charge order, spin fluctuations, and superconductivity in underdoped cuprates remain controversial.  We use neutron scattering techniques to study these phenomena in La$_{1.93}$Sr$_{0.07}$CuO$_4$, a superconductor with a transition temperature of $T_c = 20$~K.  At $T\ll T_c$, we find incommensurate spin fluctuations with a quasielastic energy spectrum and no sign of a gap within the energy range from 0.2 to 15 meV.  A weak elastic magnetic component grows below $\sim10$~K, consistent with results from local probes.    Regarding the atomic lattice, we have discovered unexpectedly strong fluctuations of the CuO$_6$ octahedra about Cu-O bonds, which are associated with inequivalent O sites within the CuO$_2$ planes.  Furthermore, we observed a weak elastic $(3\bar{3}0)$ superlattice peak that implies a reduced lattice symmetry.  The presence of inequivalent O sites rationalizes various pieces of evidence for charge stripe order in underdoped \lsco.   The coexistence of superconductivity with quasi-static spin-stripe order suggests the presence of intertwined orders; however, the rotation of the stripe orientation away from the Cu-O bonds might be connected with evidence for a finite gap at the nodal points of the superconducting gap function.
\end{abstract}
\maketitle

\section{Introduction}

The degree to which charge, spin, and superconducting orders compete or coexist is a topic of considerable current interest \cite{keim15,frad15}.  In cuprates with spatially-uniform $d$-wave superconducting order, one generally finds the absence of spin order plus a gap in the antiferromagnetic spin fluctuations, comparable in energy to the superconducting gap \cite{scal12a,esch06,yu09,ross91,chri04}.  In contrast, there are certain cuprates, such as \lbco (LBCO), that exhibit some degree of spin and charge stripe order, where the coexistence of gapless antiferromagnetic spin fluctuations and superconductivity \cite{tran08,xu14} have been interpreted as evidence for a spatially-modulated pair-density-wave superconducting order \cite{frad15,hime02,berg09b}.

In the present paper, we examine two aspects of coexisting antiferromagnetic and superconducting orders in \lsco\ (LSCO) with $x=0.07$ and superconducting transition $T_c = 20$~K.  First, we demonstrate the development of static spin order at temperatures below $T_c$, as well as the absence of any gap in the low-energy spin excitations.  Study of the elastic scattering yields a detailed characterization of the structure of the short-range spin correlations, significantly extending previous observations on a crystal with $x=0.07$ ($T_c=17$~K) \cite{fuji02c} and allowing an instructive comparison with $x=0.10$ ($T_c=29$~K) \cite{kimu99}.  The spin order appears as incommensurate peaks about the antiferromagnetic wave vector, with the incommensurate peaks rotated away from the Cu-O bond directions, as observed previously in LSCO with $x=0.12$ \cite{kimu00b} and La$_2$CuO$_{4+y}$ \cite{lee99} (but different from LBCO \cite{huck11}).  We note that this is compatible with the identification of charge-stripe order developing at 90 K, as determined by a nuclear quadrupole resonance study by Hunt {\it et al.}\ \cite{hunt99}.  The spin fluctuations have been measured well below $T_c$ and with several different incident neutron energies, providing reliable results between 0.2 and 15 meV.  The absence of a spin gap confirms and extends the results of previous studies on crystals with $x=0.07$ \cite{hira01}, 0.085 \cite{lips09}, and 0.10 \cite{lee00}.

\begin{figure}[t]
\centerline{\includegraphics[width=0.9\columnwidth]{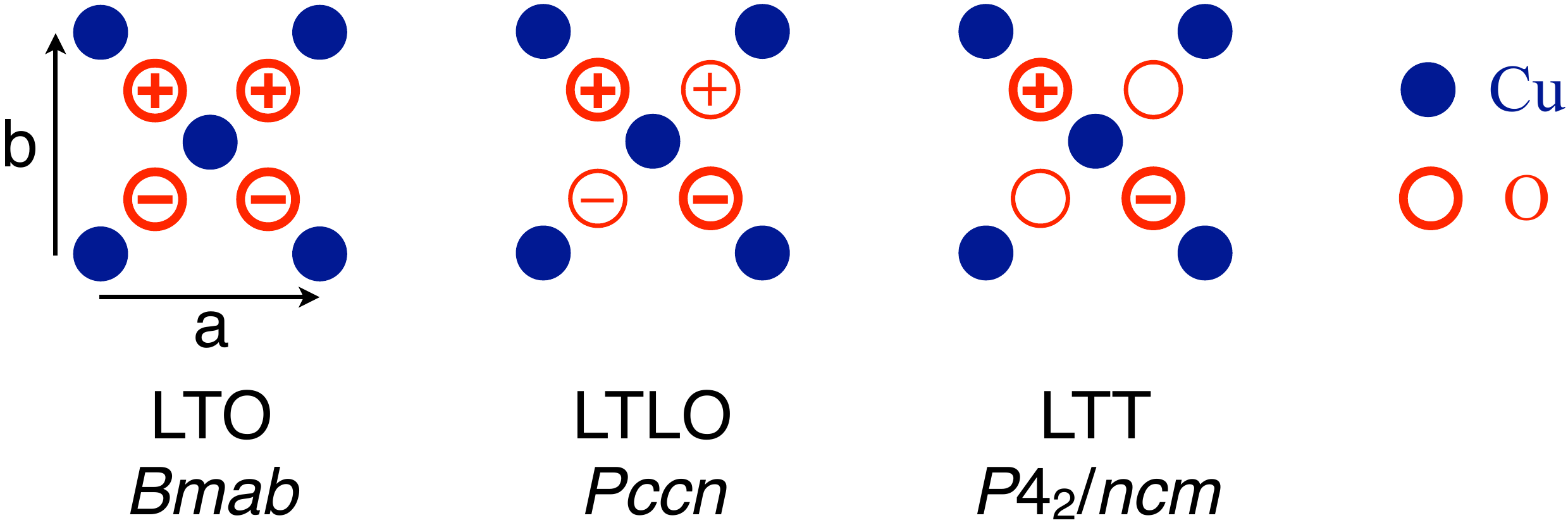}}
\caption{(color online) Diagrams indicating out-of-plane displacements of the in-plane oxygens in the low-temperature orthorhombic (LTO), low-temperature less-orthorhombic (LTLO), and low-temperature tetragonal (LTT) phases; space groups listed under acronyms. $+$ ($-$) indicates displacement above (below) plane; line thickness reflects magnitude of displacement.  There are two inequivalent in-plane O positions in LTLO and LTT, but all are equivalent in LTO.}
\label{fg:io} 
\end{figure}

We also address the nature of the stripe pinning in LSCO.  The atomic displacement patterns within the CuO$_2$ planes for the relevant crystal structures are indicated schematically in Fig.~\ref{fg:io}.  In the LTO phase, although the in-plane oxygens break $C_4$ symmetry, their positions are nevertheless symmetry related, with all belonging to the same Wyckoff position.  In LBCO, it is the presence of two inequivalent O sites within the CuO$_2$ planes of the LTT (or LTLO) phase \cite{axe89} that is associated with stripe pinning \cite{fuji04}; in both LTT and LTLO, the in-plane oxygens require two different Wyckoff positions.  It is the resulting electronic anisotropy that can pin charge stripes which tend to orient along Cu-O bonds.

Recently, the CuO$_6$ octahedral tilt fluctuations directly associated with the inequivalent O sites have been detected in LBCO by inelastic neutron scattering \cite{bozi15}.  We have now detected these fluctuations in our LSCO sample, as well.  In addition, there is a weak $(3\bar{3}0)$ superlattice peak that decreases gradually on warming to 300~K; its presence implies a lowering of the symmetry, likely to the LTLO phase, and a static inequivalence of planar O sites. The static symmetry reduction provides an explanation for previous indications of stripe pinning at relatively high temperatures in LSCO \cite{hunt99}.  In addition, the relative intensity of the tilt fluctuations is large compared to the superlattice intensity even at low temperature, indicating the existence of significant entropy associated with these tilts.

The rest of the paper is organized as follows.  The following section describes the experimental methods.  The experimental results, including the characterization of the superconducting transition of the sample and the neutron scattering measurements of spin correlations and soft phonons, are presented in Sec.~\ref{sc:results}.  We discuss these results, their relevance to the concepts of spatially-modulated superconductivity and stripe pinning, and their connection to other recent work in Sec.~\ref{sc:discussion}.  The paper is summarized in Sec.~\ref{sc:summary}.

\section{Experimental Methods}

The crystals of La$_{1.93}$Sr$_{0.07}$CuO$_4$ were grown at Brookhaven in an infrared image furnace by the traveling-solvent floating-zone method \cite{wen08,tana89}.  After growth, the diameter of the cylindrical crystal was approximately 6~mm.  The superconducting transition was determined from a measurement of the temperature dependence of the magnetization using a commercial SQUID (superconducting quantum interference device) magnetometer on a small piece of the crystal.   

For neutron scattering, two large crystals were cut from the original rod, with a total mass of 19~g.  Prior to each measurement, the two crystals were co-aligned at room temperature to yield the total effective mosaic $\lesssim 0.8^\circ$, consistent with the $\approx 0.5^\circ$ mosaic of each of the two pieces.   (For a photo of the co-mounted crystals, see inset of Fig.~\ref{fg:bragg}.)  The domain structure of the low-temperature phase was  characterized on triple-axis spectrometer HB1 at the High Flux Isotope Reactor (HFIR), Oak Ridge National Laboratory (ORNL).  The measurement was performed with a fixed final energy of 13.5~meV and collimations of $15'$-$20'$-$20'$-$240'$.   Further measurements were carried out on the SPINS cold-neutron triple-axis spectrometer at the NIST Center for Neutron Research (NCNR).   With the $c$ axis vertical, scattering wave vectors ${\bf Q}=(h,k,0)$ were accessible in the horizontal scattering plane.  Wave vectors will be expressed in units of $(2\pi/a,2\pi/b,2\pi/c)$ with the measured low-temperature lattice parameters $a=5.324$~\AA, $b=5.385$~\AA, and $c=13.2$~\AA.  Based on neutron powder diffraction measurements, the crystal structure previously has been identified as LTO phase \cite{rada94}.

Next, the sample was transferred to the Spallation Neutron Source (SNS), ORNL, where it was studied on the HYSPEC instrument (beam line 14B) \cite{hyspec15}.  In the transfer, a 2.7-g piece of crystal came loose and was removed, leaving a sample mass of 16.3 g.  The sample was mounted in a Displex closed-cycle cryostat.  Measurements were done with a fixed incident energy of either 8 or 27 meV and a chopper frequency of 300 Hz.   The graphite-crystal array in the incident beam was set for vertical focusing. The energy resolution (full-width half maximum) near elastic scattering is $\sim 0.25$~meV for $E_i=8$~meV and $\sim 1$~meV for $E_i=27$~meV. For a typical measurement, the detector tank was placed at a mean scattering angle of either 33$^\circ$ or 80$^\circ$, and then measurements were collected for a series of sample orientations, involving rotations about the vertical axis in steps of $1^\circ$.  (The position-sensitive detector collects neutrons for scattering angles of $\pm30^\circ$ in the horizontal and $\pm7.5^\circ$ in the vertical direction.)  From such a set of scans, a four-dimensional (4D) data set was created and analyzed with the MANTID~\cite{mantid14} and HORACE~\cite{Horace} software packages.  Slices of data corresponding to particular planes in energy and wave-vector space can then be plotted from the larger data set.  For reference, the beam power was $\sim0.85$~MW, and the counting time for a single 4D data set was roughly 6 h.

\section{Results}
\label{sc:results}

\subsection{Sample characterization}

\begin{figure}[t]
\centerline{\includegraphics[width=\columnwidth]{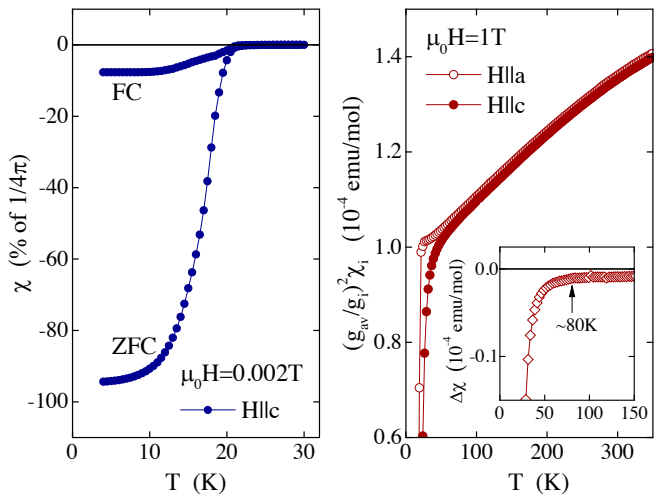}}
\caption{(color online)  (a) Bulk susceptibility of the LSCO $x=0.07$ sample measured with a 20 G magnetic field applied along the $c$ axis both for field cooling (FC) and zero-field cooling (ZFC) conditions.  (b) Comparison of magnetic susceptibility measured in a 1-T field applied either along an in-plane Cu-O direction (open circles, labelled $H \| a$) or along the $c$ axis (filled circles) after scaling by the anisotropic $g$ factors taken from \cite{huck08}.  Inset shows the difference between these scaled susceptibilities, indicating that diamagnetism within the CuO$_2$ planes sets in below $\sim 80$~K.  (To convert the susceptibility to SI units of H\,m$^2$\,/\,mol, multiply emu/mol by $(4\pi)^2\times10^{-13}$.)}
\label{fg:chi} 
\end{figure}

The bulk magnetic susceptibility measured at low field on a piece of the LSCO $x=0.07$ sample is shown in Fig.~\ref{fg:chi}(a).  As one can see, the bulk transition is at 20~K.  The data have been corrected for the demagnetizing factor, and within the 10\%-uncertainty of that correction, the superconducting shielding fraction is consistent with 100\%.  Measurements on pieces taken from different positions along the crystal rod indicate a variation of $T_c$ of $<1$~K.

There is also some weak anisotropic diamagnetism above the bulk transition.  To characterize that, the susceptibility was measured with a field of 1~T applied along the $c$ axis and along an in-plane Cu-O bond direction.  Assuming isotropic spin susceptibility at high temperature ($>150$~K) and using the anisotropic $g$ factors determined in a previous study of LBCO \cite{huck08}, we obtain the corrected spin susceptibilities shown in Fig.~\ref{fg:chi}(b).  From the difference curve plotted in the inset, one can see that the diamagnetism of the CuO$_2$ planes becomes significant below $\sim80$~K.   This is consistent with previous results on LBCO, especially with $x=0.095$ \cite{wen12a}, and with studies of the onset of superconducting fluctuations in LSCO by torque magnetometry \cite{li10} and Nernst effect \cite{wang01}.

\begin{figure}[t]
\centerline{\includegraphics[width=\columnwidth]{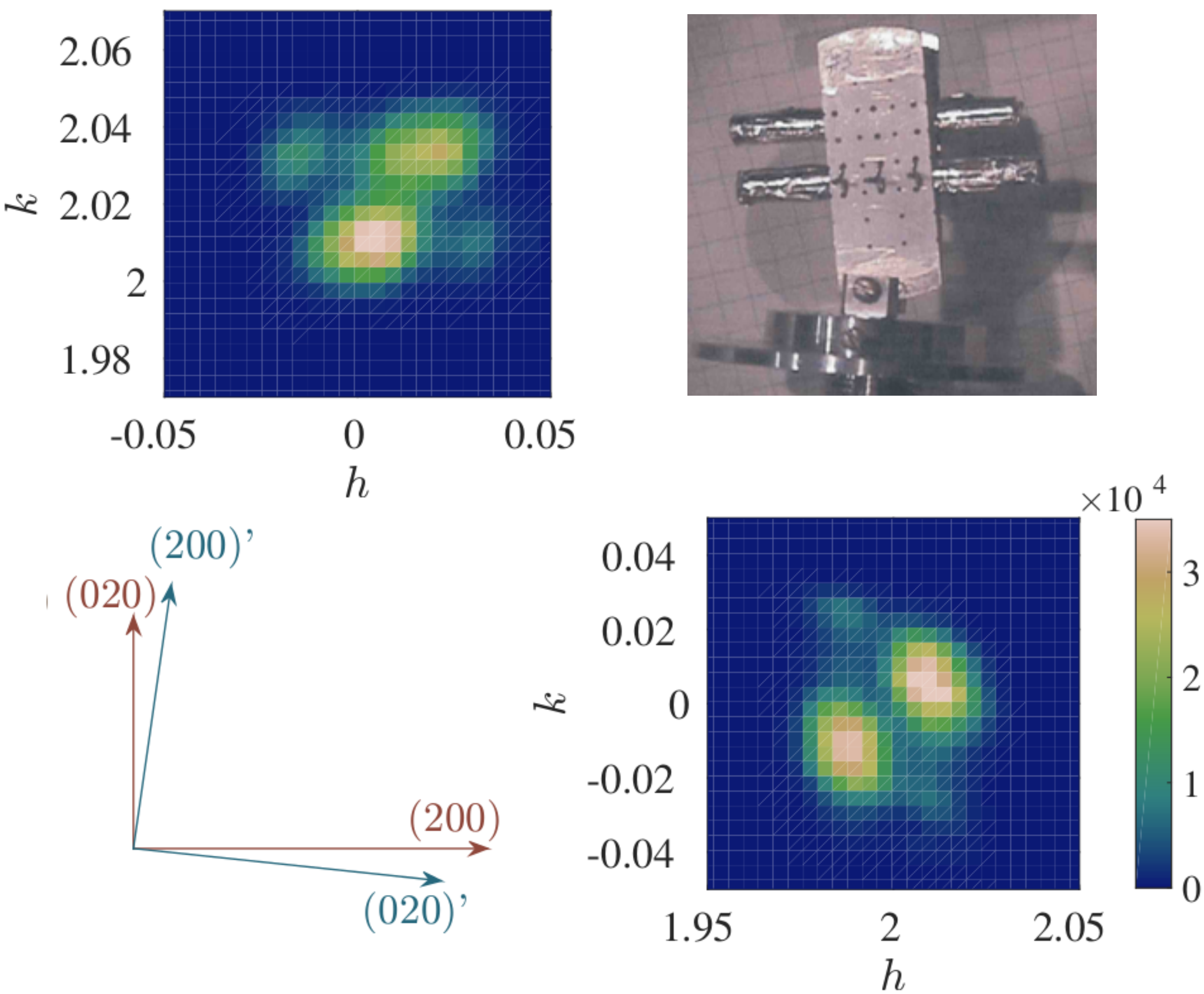}}
\caption{(color online)  Arrows indicate schematically the orientations of (200) and (020) fundamental Bragg peaks of the two main twin domains.  Image plots show the results of mesh scans performed around these Bragg peaks.  Upper right inset is a photo of the co-mounted crystals, each wrapped in Al foil.}
\label{fg:bragg} 
\end{figure}

The twin-domain structure of the orthorhombic phase at $T=4.2(1)$~K is presented in Fig.~\ref{fg:bragg}, which shows mesh scans covering the positions of the (200) and (020) Bragg peaks.  As indicated schematically by the arrows, we have two main twin domains.  Least squares fitting indicates that the intensity ratio of the (200) domain to the $(020)'$ is 1.08(6).

\subsection{Magnetic scattering}

\subsubsection{Elastic}
\label{mag_el}

\begin{figure}[b]
\centerline{\includegraphics[width=\columnwidth]{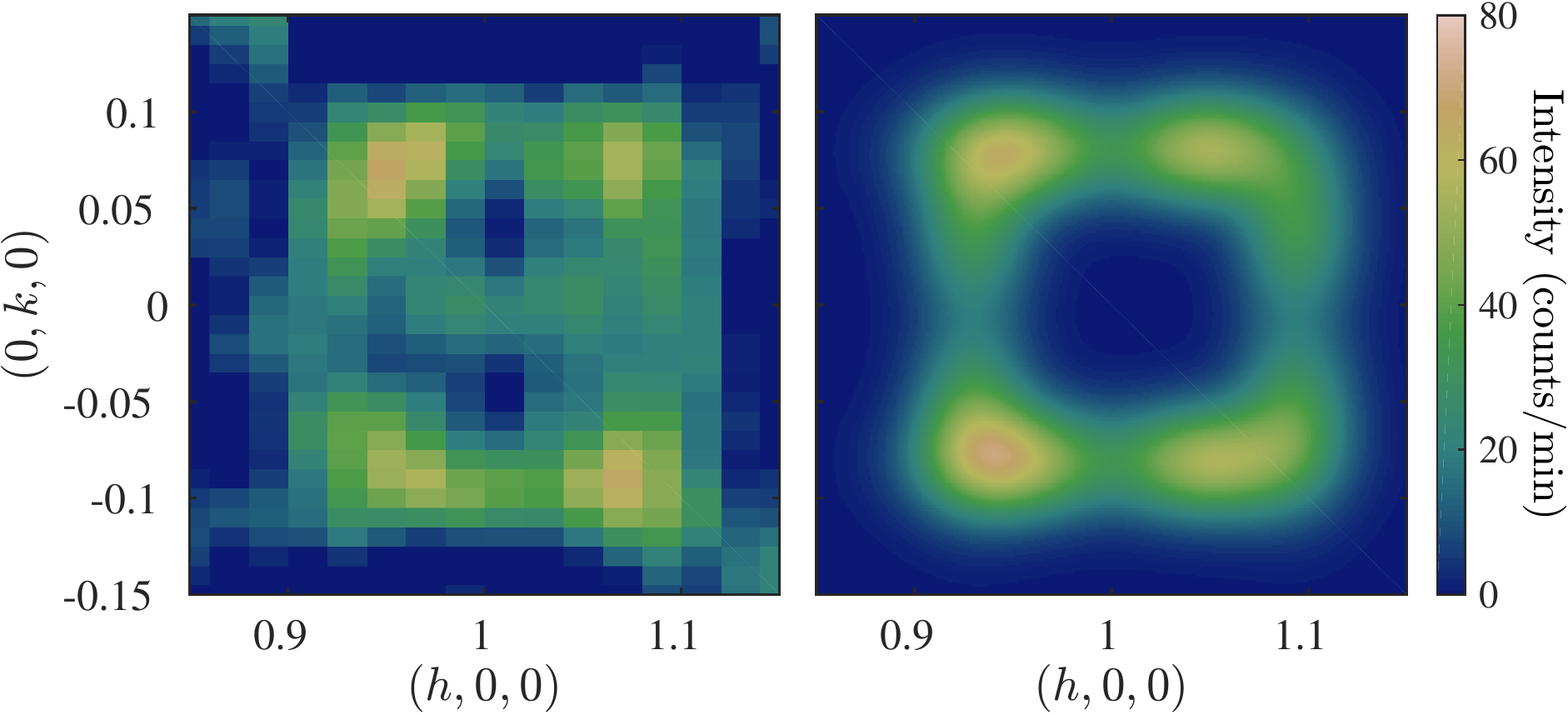}}
\caption{(color online)  Elastic intensities around the antiferromagnetic wave vector at 1.5~K, (a) measured (with background subtracted) and (b) fit, as described in text.}
\label{fg:qxqy} 
\end{figure}

\begin{figure}[b]
\centerline{\includegraphics[width=0.9\columnwidth]{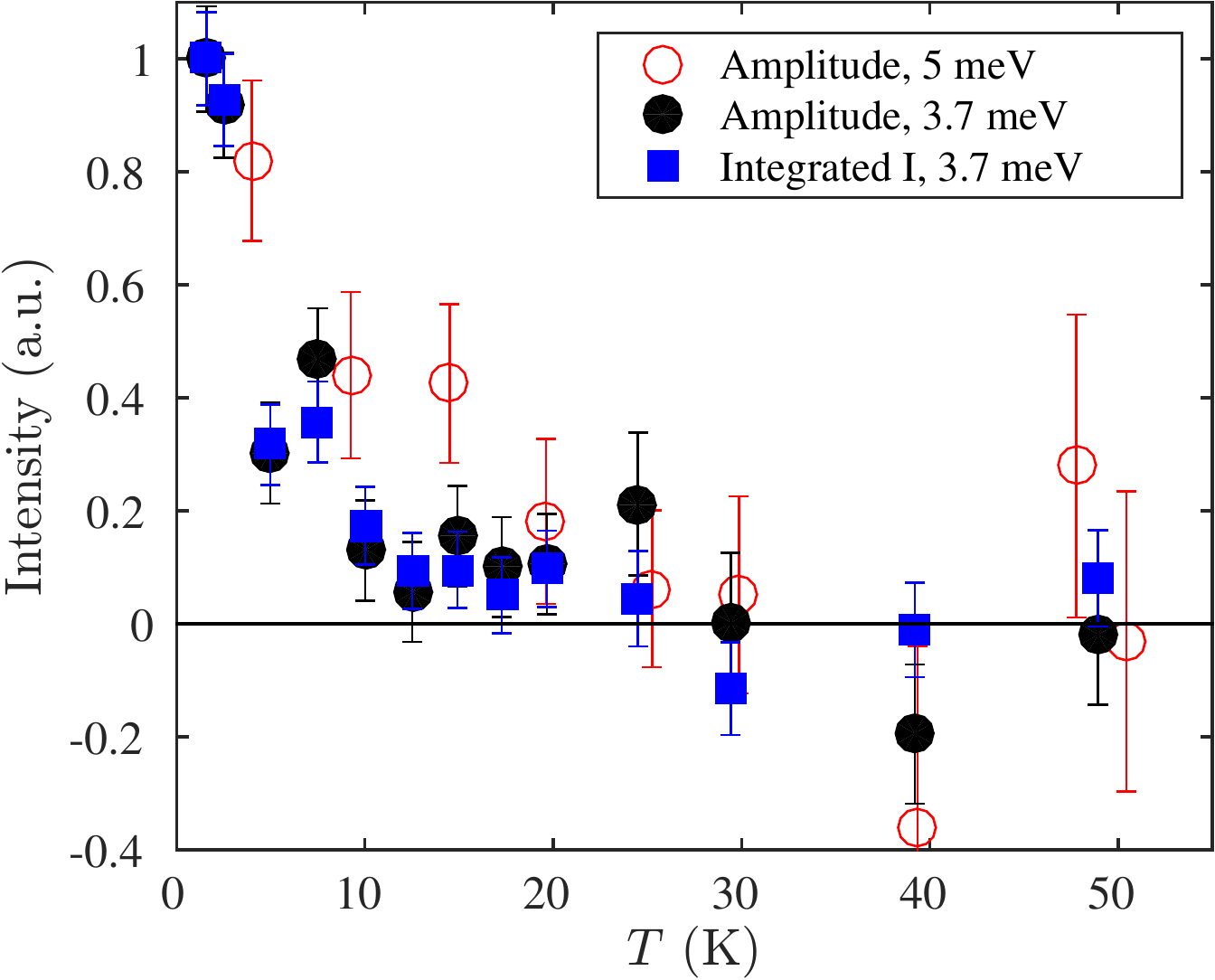}}
\caption{(color online)  Temperature dependence of the intensity of elastic magnetic scattering, as described in the text.  Filled symbols represent results obtained with $E_f=3.7$~meV; (blue) squares for integrated intensity, (black) circles for peak amplitude; results normalized to 1 at the lowest temperature.  Open circles (red) indicate peak amplitude obtained with $E_f=5$~meV.    Error bars here and elsewhere correspond to $1\sigma$ determined from counting statistics.}
\label{fg:tdep} 
\end{figure}

\begin{figure*}[t]
\centerline{\includegraphics[width=1.95\columnwidth]{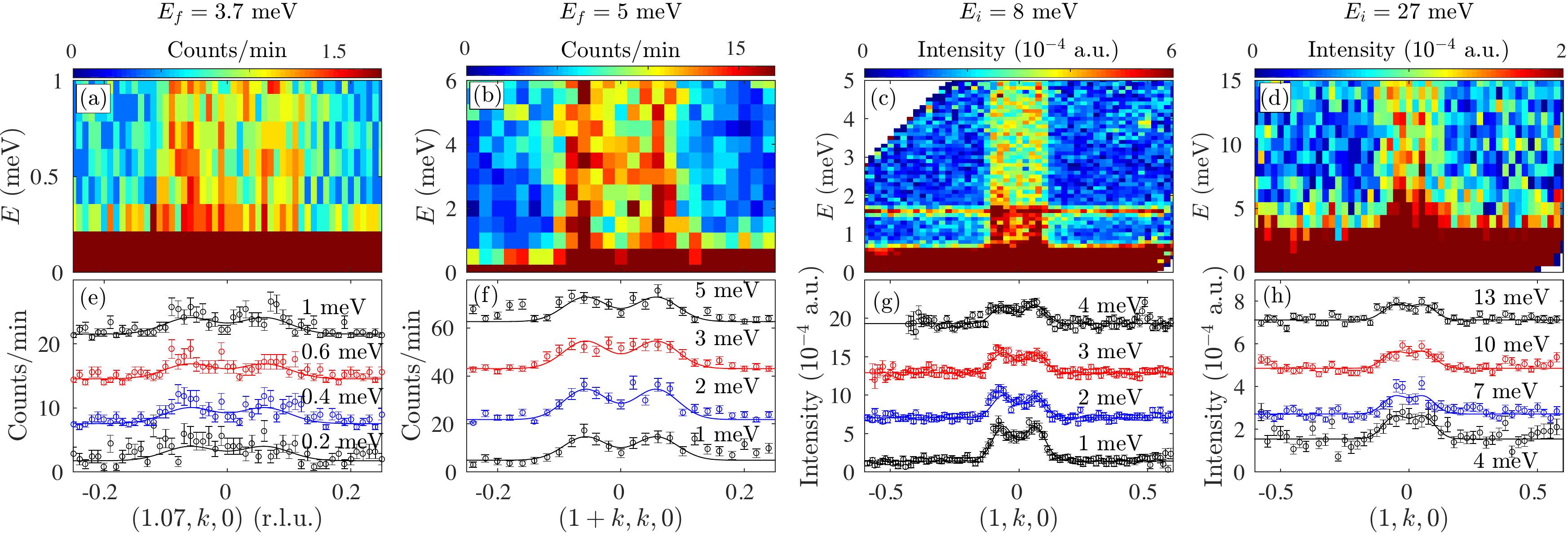}}
\caption{(color online)  Image plots of intensity vs.\ $E$ and {\bf Q} for (a) SPINS with $E_f=3.7$~meV, (b) SPINS with $E_f=5$~meV, (c) HYSPEC with $E_i=8$~meV, (d) HYSPEC with $E_i=27$~meV.  Representative scans for (e) $E_f=3.7$~meV, (f) $E_f=5$~meV.  Cuts integrated over $0.9\le h\le1.1$ and $-0.25\le l\le0.25$ for (g) $E_i=8$~meV, (h) $E_i=27$~meV.  Lines in (e)-(h) represent fits of a pair of symmetric gaussian peaks on a {\bf Q}-independent background.  Scans are vertically offset for clarity.  In (c), there is a spurious feature (of undetermined origin) at $E\approx1.7$~meV.}
\label{fg:Edep} 
\end{figure*}

We begin with the characterization of the elastic magnetic scattering about ${\bf Q}_{\rm AF}=(1,0,0)$.  A mesh of scans was measured at SPINS with $E_f=5$~meV for a temperature of $T=1.5(1)$~K, and the low $Q$ part of the mesh was also obtained at 30~K.  As no magnetic signal was detected at 30~K, that data set provided a useful measurement of the background.  The 1.5-K data were fitted with four gaussian peaks from each twin domain, at ${\bf Q}_{\rm AF}\pm {\bf q}_\delta^\pm$ with ${\bf Q}_{\rm AF}=(1,0,0)$ and at ${\bf Q}'_{\rm AF}\pm {\bf q}_\delta^\pm$ with ${\bf Q}'_{\rm AF}=(0,1,0)'$, plus a gaussian in $h$ and $k$ to describe the background; the same background parameters were also fit simultaneously to the 30-K data.  The 1.5-K data, after background subtraction, are shown in Fig.~\ref{fg:qxqy}(a); the fitted peaks are displayed in panel (b).  The peak splitting ${\bf q}_\delta^\pm = (\delta h,\pm\delta k,0)$ corresponds to $\delta h = 0.048(2)$ and $\delta k = 0.081(1)$, where the number in parentheses is the uncertainty of the last digit of the preceding number.   [For comparison, fitting gaussian peaks while assuming a single domain yields positions $\delta h = 0.059(2)$ and $\delta k = 0.074(2)$.]

The rotation of the peaks away from $\delta h = \delta k$ (that is, a rotation of the modulation away from the Cu-O bond direction) is qualitatively consistent with the peak rotation seen previously in \lsco\ with $x=0.12$ \cite{kimu00,tham14} and in \lcod\ \cite{lee99}, although the magnitude of the rotation in the present case, $14(1)^\circ$, is considerably larger.  [For the single-domain fit, the rotation would be $7(2)^\circ$.]  The intensity contribution from the (0,1,0) domain is $\lesssim60$\%\ of that from the (1,0,0), which is roughly consistent with the intensity anisotropy observed in the spin-glass regime for LSCO with $x=0.05$ \cite{waki00}.  It suggests a tendency for spins to align along $b$ (but with disorder), which is also indicated by the anisotropic magnetic susceptibility measured on detwinned crystals of $x=0.03$ \cite{lavr01}.   In our fitting, the gaussian widths were allowed to be anisotropic along the $h$ and $k$ directions, resulting in $\sigma_h=0.045(3)$~rlu and $\sigma_k=0.028(2)$~rlu.

The temperature dependence of the elastic peaks was probed in two ways.  With $E_f=3.7$~meV, a scan was measured along ${\bf Q}=(1.07,k,0)$ at each temperature.  The intensity was fit with a pair of gaussians plus constant background.  Both the integrated intensities and the peak amplitudes are plotted in Fig.~\ref{fg:tdep}.  With $E_f=5$~meV, the intensity was measured at a single position, $(1.07,-0.07,0)$, as a function of temperature, and the measurements have been averaged over bins of 5-K width to improve statistics; background was obtained by assuming that no signal was present above 25 K.  The resulting background-subtracted amplitude is also presented in Fig.~\ref{fg:tdep}.  

While the statistical error bars are substantial, there appears to be a clearly distinguishable difference in the temperature dependence for the two sets of measurements.  The onset temperature is correlated with the energy resolution for the measurement.  The energy resolution is slightly coarser for $E_f=5$~meV, and the onset of intensity appears to be close to $T_c$.  For the higher-resolution measurements with $E_f=3.7$~meV, the main growth in intensity occurs below $\sim 10$~K.  Going to yet higher energy/time resolution, muon spin rotation studies indicate the onset of a finite hyperfine field at $\sim4$~K \cite{weid89,nied98}.  The resolution-dependence of the onset temperature suggests the gradual freezing of slowly fluctuating spin correlations, as found in the spin-glass phase of \lsco\ for $x\lesssim0.05$ \cite{waki99,bao07}.

\subsubsection{Inelastic}

\begin{figure}[b]
\centerline{\includegraphics[width=0.7\columnwidth]{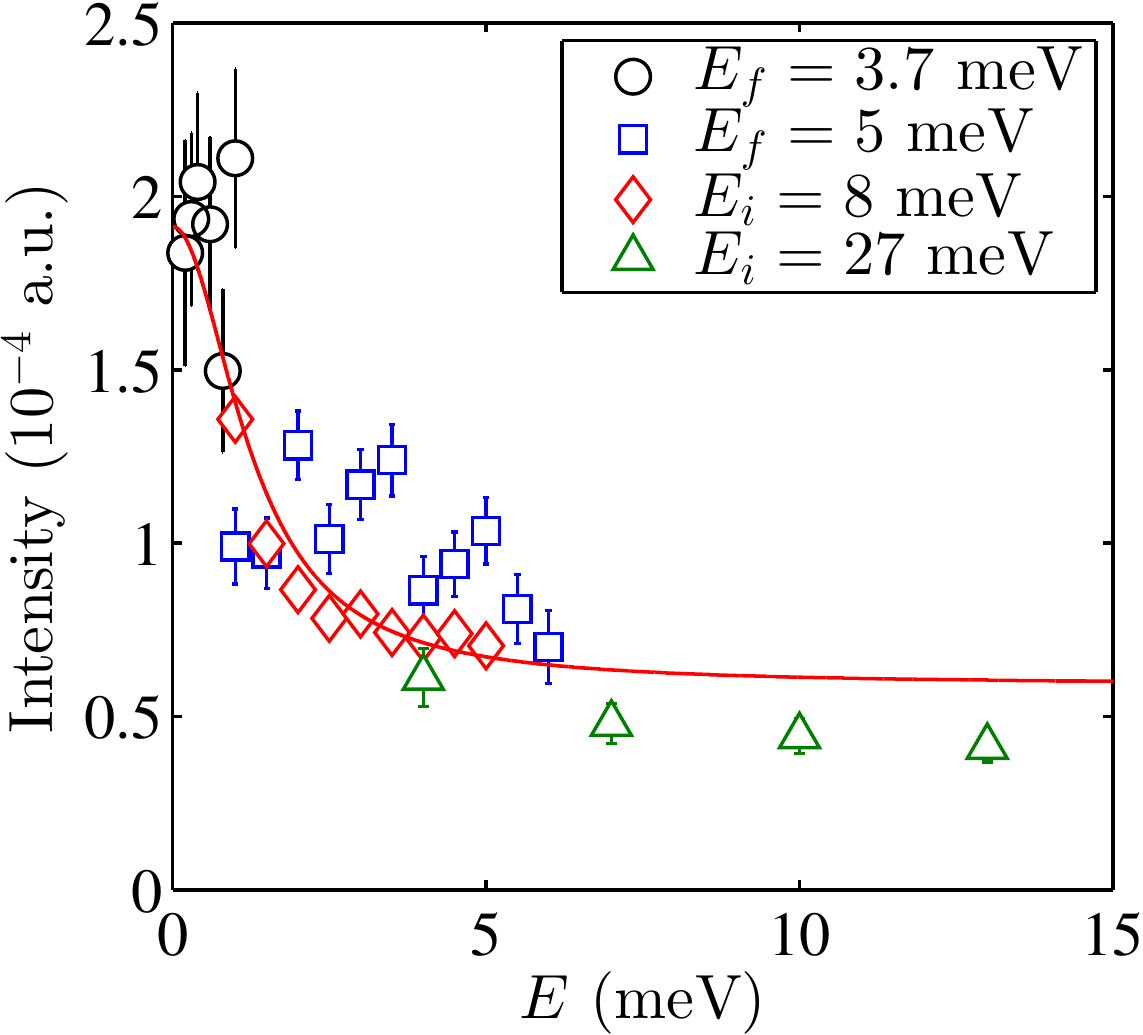}}
\caption{(color online)  Compilation of the {\bf Q}-integrated magnetic intensity (normalized as discussed in the text) as a function of excitation energy.  The line through the data is a fit to a constant plus a lorentzian centered at $E=0$; the half-width at half-maximum of the lorentzian is $1.3(2)$~meV.}
\label{fg:IvE} 
\end{figure}

For low-energy spin fluctuations, the intensity has an incommensurate pattern similar to that of the elastic signal in Fig.~\ref{fg:qxqy}(a), though without the peak rotation and with somewhat greater peak widths, as originally demonstrated by Cheong {\it et al.} \cite{cheo91}. 
Here we focus on characterizing the frequency dependency of the magnetic signal at base temperature ($\lesssim5\ \mbox{\rm K} \ll T_c$) for the range $0.2\mbox{\ meV}\le\hbar\omega\le 15$~meV.  Figure~\ref{fg:Edep} shows representative scans (from SPINS) and cuts (from HYSPEC data) for constant energies as a function of ${\bf Q}$, as well as images of the intensity as a function of $E=\hbar\omega$ and ${\bf Q}$.  The lines through the data points in Fig.~\ref{fg:Edep}(e)-(f) indicate fitted gaussian peaks used to evaluate the integrated intensity.    For each measurement condition, a distinct range of excitation energies is covered; in each case, there is substantial intensity at the lowest resolvable energies.   For the gaussian fits, the peaks are symmetric in $k$, with peak center and gaussian width  allowed to vary independently for each set of data, but constrained to be independent of excitation energy.  All of the results are consistent with a peak center of $\delta k=0.060(4)$~rlu and gaussian width $\sigma_k = 0.041(5)$ rlu.  The magnitude of incommensurability for the inelastic scattering is about 10\%\ smaller than that of the elastic signal; this is qualitatively consistent with the results for LBCO with $x=0.095$ \cite{xu14}.

\begin{figure*}[t]
\centerline{\includegraphics[width=1.5\columnwidth]{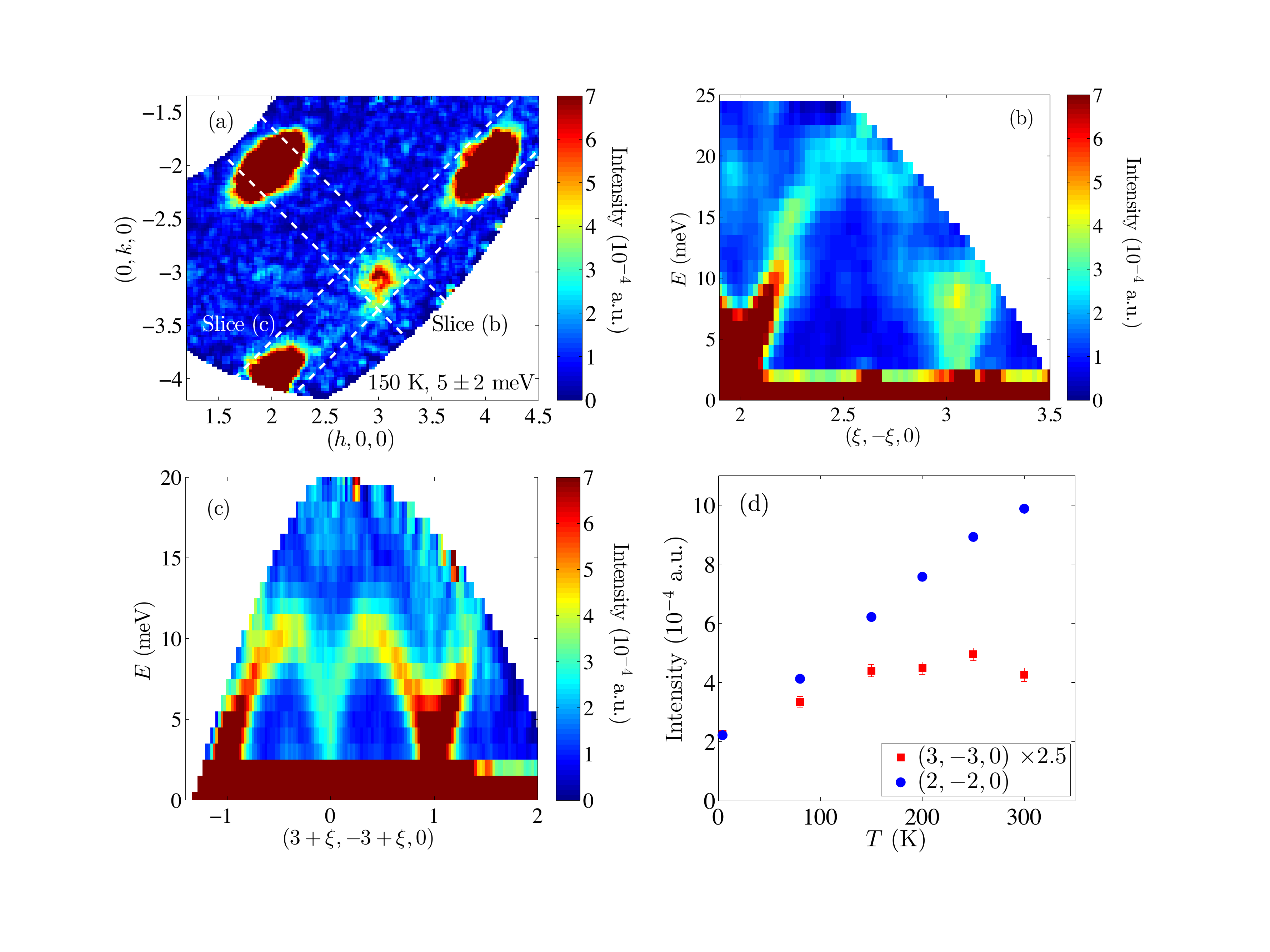}}
\caption{(color online)  Measurements of phonons in \lsco\ with $x=0.07$ at $T=150$~K.  (a) Constant-energy slice ($\hbar\omega = 5\pm2$~meV with $L$ integrated from $-0.25$ to 0.25), showing strong acoustic phonons about the $(2,-2,0)$, $(2,-4,0)$, and $(4,-2,0)$ Bragg points, and soft tilt fluctuations at $(3,-3,0)$.  Dashed lines indicate the directions and widths of the slices through $(3,-3,0)$ shown in (b) (longitudinal direction) and (c) (transverse direction).  (d) Intensity of the phonon signal at $(3,-3,0)$ integrated from 3 to 7~meV [as in (a)] and multiplied by 2.5 (red squares) and at $(2,-2,0)$ (blue circles), corrected for background measured in between these positions.}
\label{fg:phonons} 
\end{figure*}

The integrated intensities obtained from the fits to the spin-fluctuation scattering shown in Fig.~\ref{fg:Edep} are presented in Fig.~\ref{fg:IvE}.  Here, each data set has been normalized to a measurement of the elastic incoherent scattering, which is dominated by the incoherent nuclear scattering from the sample; the spread in the results is consistent with the uncertainty in the normalization.  We observe a peak at $E\approx0$, with the intensity falling off to a constant level for $E\gtrsim5$~meV.  The line through the data corresponds to a constant plus a lorentzian centered at $E=0$ with a half-width of 1.3(2) meV.  The absence of a spin gap in the superconducting state is consistent with a recent study of \lbco\ with $x=0.095$ \cite{xu14}.

\subsection{Lattice response}

\subsubsection{Phonons}

Soft phonons corresponding to octahedral tilts that break the equivalence of in-plane O sites were recently identified and studied in \lbco\ with $x=0.125$ \cite{bozi15}; ordering of these displacements leads to a reduced lattice symmetry that can pin charge stripes \cite{tran95a,fuji04}.  Given the evidence for charge stripe order in \lsco\ \cite{hunt99,chri14,crof14,tham14}, we decided to test for such soft tilts.  We did this by measuring about $(3,-3,0)$, the same position studied in \lbco\ and a wave vector at which elastic scattering is forbidden in the LTO phase.

Representative results, obtained at $T=150$~K, are presented in Fig.~\ref{fg:phonons}(a)-(c).  Panel (a) shows inelastic signal, integrated between 3 and 7 meV, and plotted vs.\ wave vector.  The elliptical features at $(2,-4,0)$, $(2,-2,0)$, and $(4,-2,0)$ are acoustic phonons about fundamental Bragg peaks, while the intensity at $(3,-3,0)$ corresponds to soft tilt modes.  Panels (b) and (c) show the dispersion through $(3,-3,0)$ along longitudinal and transverse directions, respectively.  One can see a continuous connection to transverse acoustic phonons in (c).  For the longitudinal fluctuations in (b), the intensity appears to be interrupted by an anti-crossing with an unseen mode at $\sim12$ meV that also impacts the longitudinal acoustic mode dispersing from $(2,-2,0)$; nevertheless, the upper part of the the dispersion can be seen between 15 and 20 meV.  If the signal at $(3,-3,0)$ were a soft mode, we would expect to see an energy gap; however, there is no sign of one.  To the extent a gap might be present, it would have to be highly over-damped, which does not seem to agree with the dispersion clearly seen in Fig.~\ref{fg:phonons}(c). 

We repeated these measurements at several temperatures.   To compare the temperature dependence of these data, we integrated the signal centered on $(3,-3,0)$ from 3 to 7 meV; the results, after background subtraction, are indicated by red squares in Fig.~\ref{fg:phonons}(d).  For comparison, the blue circles indicate a similar integral over acoustic phonons about the $(2,-2,0)$ Bragg peak.   The intensity at the ``forbidden'' position shows little change with temperature between 150 and 300~K; this is similar to the constant intensity found for the same mode within the LTO phase of \lbco\ with $x=0.125$ \cite{bozi15}.  On the other hand, there is some decrease in intensity on cooling to 80 and 5~K.   The main observation here is that the two temperature dependencies are different: that of the lattice phonon at the fundamental Bragg reflection is consistent with the detailed balance and therefore a $T$-independent imaginary part of the dynamical susceptibility, $\chi''$, while the tilt mode indicates temperature-dependent $\chi''$, with the characteristic temperature of $\sim150$~ K. 

\subsubsection{Superlattice peak}

\begin{figure}[b]
\centerline{\includegraphics[width=1\columnwidth]{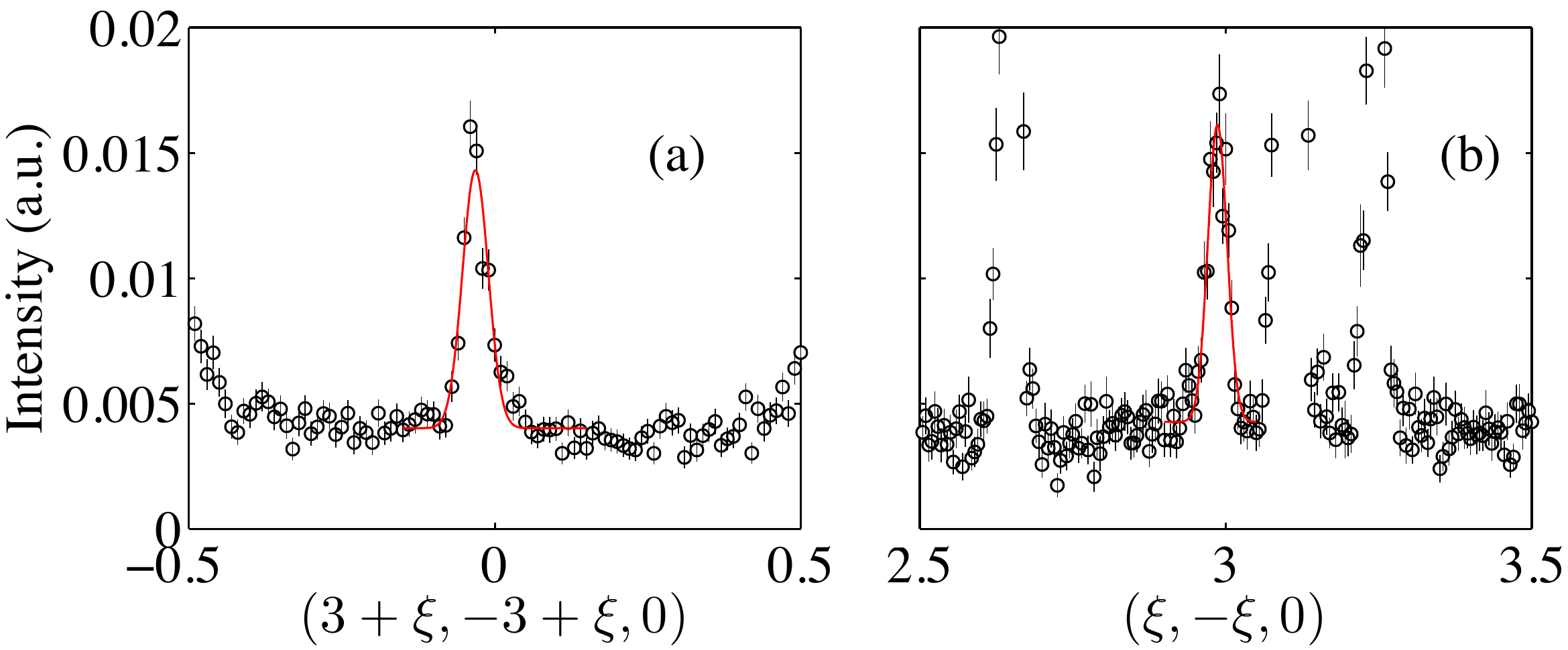}}
\caption{(color online) Elastic scattering ($\hbar\omega=0\pm0.5$~meV) about the $(3,-3,0)$ position along the (a) transverse and (b) longitudinal directions.  In each case, the red curve is a gaussian fit to the $(3,-3,0)$ peak.  The extra peaks in (b) are powder diffraction peaks from the Al sample holder. }
\label{fg:el} 
\end{figure}

\begin{figure}[t]
\centerline{\includegraphics[width=0.8\columnwidth]{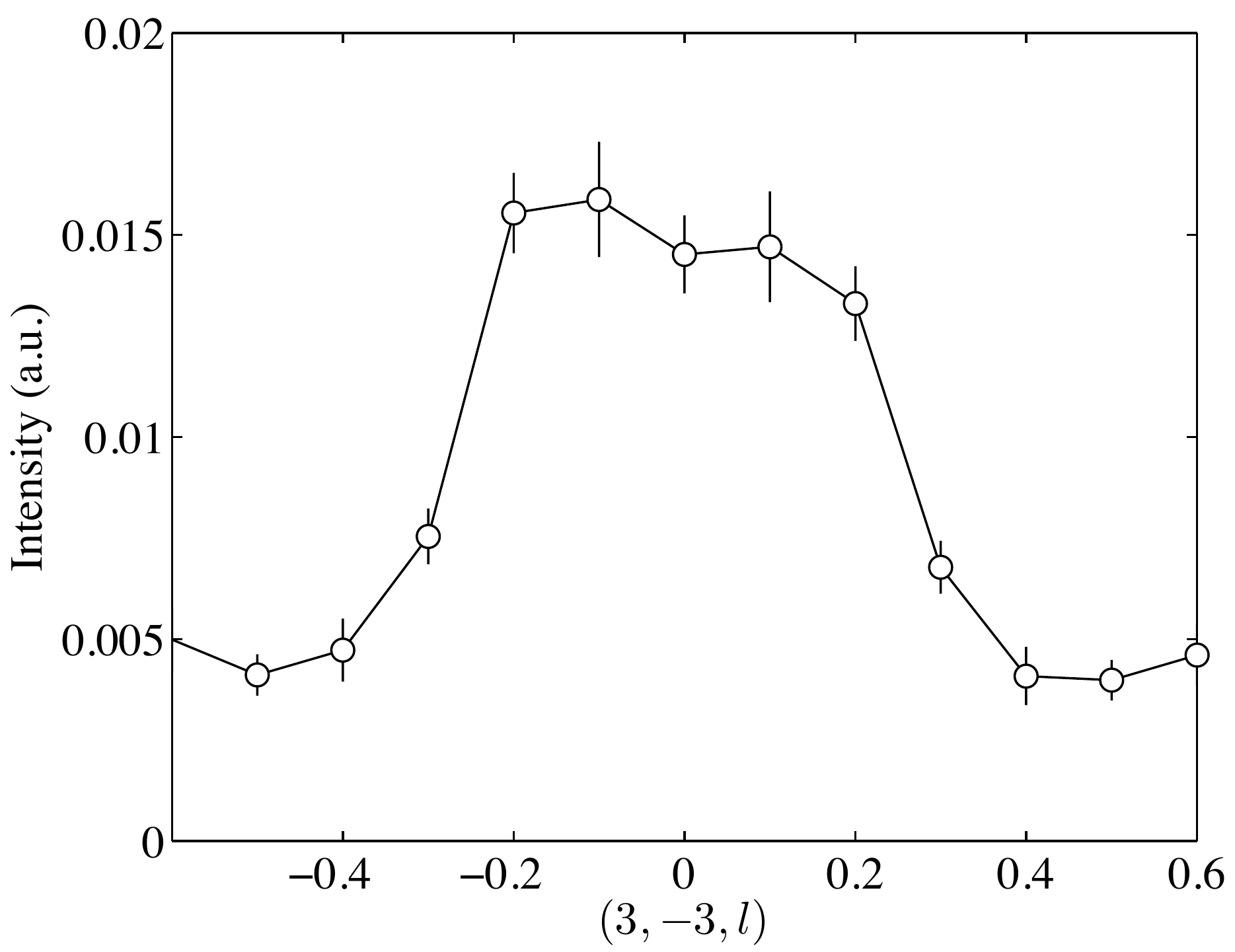}}
\caption{Elastic scattering along ${\bf Q}=(3,-3,l)$.  Based on a comparison with a cut along $(2,-2,l)$, we conclude that the peak shape and width are due to the vertical focusing and the sample geometry.  }
\label{fg:el2} 
\end{figure}

\begin{figure}[b]
\centerline{\includegraphics[width=0.7\columnwidth]{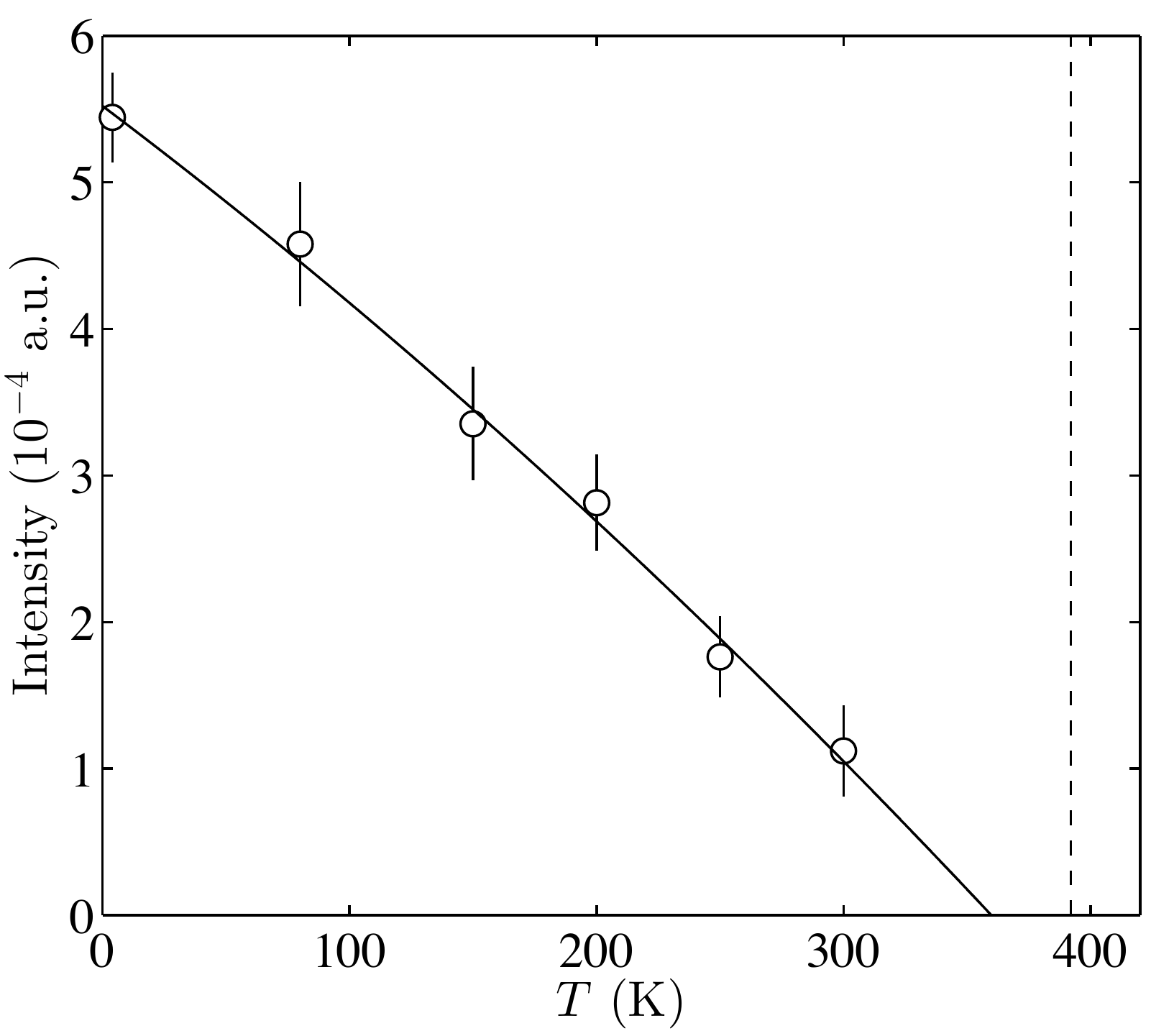}}
\caption{Integrated intensity of the $(3,-3,0)$ superlattice peak as a function of temperature.  The vertical dashed line indicates the transition to the HTT phase based on an interpolation formula \cite{taka94}.  }
\label{fg:el_T} 
\end{figure}

Given the strong LTT fluctuations, we have also looked at the elastic scattering at the $(3,-3,0)$ superlattice position.  In-plane cuts along transverse and longitudinal directions are shown in Fig.~\ref{fg:el}; the out-of-plane cut is shown in Fig.~\ref{fg:el2}.   We find a peak that appears to have a resolution-limited width in all three directions.  The temperature-dependence of the integrated intensity (evaluated from the transverse cuts) is shown in Fig.~\ref{fg:el_T}.  It decays only gradually with temperature, and heads to zero close to the estimated temperature for the transition to the high-temperature tetragonal (HTT) phase \cite{taka94}.  Given that the same superlattice peak was found to go to zero in the LTO phase of LBCO $x=0.125$ under similar measurement conditions \cite{bozi15}, it seems unlikely that the present peak could have a spurious cause, such as double scattering.  Hence, it appears that the symmetry of the crystal structure is lower than that suggested by the analysis of neutron powder diffraction data \cite{rada94}.  As the average structure has a significant orthorhombic strain, the likely structure is the low-temperature less orthorhombic (LTLO; $Pccn$) phase.  

It is of interest to compare the intensity of the superlattice $(3,-3,0)$ peak to a fundamental Bragg peak such as $(2,-2,0)$.   At 5~K, we find an intensity ratio of $7\times10^{-4}$; this is an upper limit, as the $(2,-2,0)$ intensity almost certainly saturated the detector, as well as suffering secondary extinction from our large crystals.  Such an intensity ratio is definitely outside of the dynamic range of a powder-diffraction measurement.  We can also compare it with the ratio of low-energy phonon intensities, as evaluated in Fig.~\ref{fg:phonons}(d); the acoustic phonon intensity at wave vectors close to the Bragg peak should be proportional to the elastic peak intensity.  From the figure we find an intensity ratio at 5~K of 0.4, three orders of magnitude greater than the elastic intensity ratio.  This indicates that there is substantial entropy remaining in these tilt fluctuations at low temperature.

\section{Discussion}
\label{sc:discussion}

\subsection{Glassy spin-stripe order}

The elastic magnetic intensity data in  Fig.~\ref{fg:tdep} suggest an onset temperature in the range of 10--20~K.  A considerably lower transition temperature is indicated by nuclear magnetic resonance (NMR) and muon spin rotation ($\mu$SR) techniques, which are sensitive to much smaller time scales.  $^{139}$La NMR and nuclear quadrupole resonance (NQR) studies of LSCO indicate a spin freezing temperature of 5~K for $x=0.06$ \cite{juli99} and a transition below 4.2~K for $x=0.07$ \cite{baek12}.  This frequency dependence of the freezing transition indicates glassy behavior.  In fact, the transition temperature increases smoothly on reducing the doping into the non-superconducting regime $0.02 \lesssim x\lesssim 0.055$ \cite{nied98,juli03}, where spin-glass behavior was originally identified in measurements of the bulk magnetization \cite{chou95,waki00b}.

Even before the glassy response was experimentally detected, it was proposed that the doping of holes into antiferromagnetic CuO$_2$ layers would lead to a ``cluster'' spin glass due to frustrated electronic phase separation \cite{emer93}.  Indeed, $\mu$SR measurements at $T<1$~K have identified an average local hyperfine field of finite magnitude for $0\le x<0.09$ \cite{nied98}, especially including $x=0.07$ \cite{weid89}, consistent with local regions of antiferromagnetic order.  The finite elastic peak width that we have observed indicates high sensitivity to quenched disorder; note that our doping level of 7\%\ is far below the percolation limit for magnetic dilution \cite{vajk02}.  In the context of coupled spin and charge stripe order \cite{tran95a,zach98}, it was proposed that the charge stripes would be the component that would couple strongly to electronic disorder, resulting in a stripe glass \cite{kive00}.  The generally strong sensitivity of charge-ordered phases to disorder in quasi-2D layered materials, such as cuprates, is also expected on the basis of the low-energy mapping of the interactions responsible for the ordering onto an effective random-field Ising model \cite{zach03}.  Neutron scattering measurements eventually demonstrated the presence of incommensurate elastic magnetic scattering throughout the spin-glass regime \cite{fuji02c}, consistent with a stripe glass phase.  (An incommensurate spin-spiral phase has also been proposed \cite{sush05}; however, the nearly uniaxial spin anisotropy that we observe is not consistent with a spin-spiral state.)

While the absence of long-range stripe order in the presence of disorder is inevitable in two dimensions, a new possible glassy state, the spin-density-wave (SDW) glass, has recently been proposed \cite{mros15}.  In the SDW glass, the spins maintain a preferred direction over long distances (spin nematic order).  It is the proliferation of single dislocations of the charge stripes \cite{zaan01} that would normally destroy such order.  It is presumed that, for sufficiently weak disorder, a regime may exist in which double dislocations proliferate but single dislocations are suppressed.

Could our sample exhibit SDW-glass behavior?  The fact that we observed the rotation of the incommensurate magnetic peaks away from the Cu-O directions in the presence of structural twin domains is a consequence of the tendency for the spins to align along [010], as discussed in Sec.~\ref{mag_el}.   Of course, this is essentially the same orientation of the spins found in antiferromagnetic La$_2$CuO$_4$, where the spin orientation is determined by deviations from pure Heisenberg spin coupling due to weak exchange anisotropy resulting from spin-orbit coupling effects \cite{kast98}.   Hence, while we do see a tendency towards spin-nematic order (and away from a fully random distribution of spin orientations), it is not clear whether this is direct evidence for the SDW-glass state.

\subsection{Pair-density-wave glass}

We have observed quasi-static incommensurate antiferromagnetism, with gapless magnetic excitations, at temperatures well below the superconducting transition.  These results are similar to those found in La$_{1.905}$Ba$_{0.095}$CuO$_4$, where weak charge and spin stripe order \cite{wen12} and gapless spin excitations \cite{xu14} coexist with superconductivity.    Gapless incommensurate spin fluctuations have been seen previously in LSCO with $x\lesssim0.13$ \cite{lee00,chan07,lips09,kofu09a}, although this behavior has not always been recognized as intrinsic for measurements without an applied magnetic field \cite{kofu09a}.

As noted elsewhere \cite{xu14}, the coexistence of quasi-static spin stripes with superconductivity is inconsistent with expectations for a uniform $d$-wave superconductor \cite{scal12a,esch06} but is compatible with a spatially-modulated superconducting order parameter, as with the proposed pair-density-wave state (PDW) \cite{hime02,berg07,corb14,frad15}.   The enhancement of spin-stripe order in underdoped \lsco\ by application of a $c$-axis magnetic field has been observed previously \cite{lake02}; such a field also leads to decoupling of the superconducting layers \cite{scha10,scha10b}, the effect that originally motivated the concept of the PDW superconductor \cite{hime02,berg07}.

A concern with the PDW state has been that it is very sensitive to disorder \cite{berg09b}.  A distribution of single dislocations can destroy the long-range phase order.  It happens that the SDW glass theory also applies to a PDW glass \cite{mros15}; in the latter case, it is proposed that superconducting order may survive in the form of a $4e$ nematic phase \cite{berg09c}.  (The transformation from PDW order to $4e$ nematic superconductor due to thermal fluctuations is discussed in \cite{barc11}.)

Another potential challenge to obtaining superconducting order from PDW correlations is the frustration of the interlayer Josephson coupling, as occurs in he case of LBCO with $x=0.125$ \cite{li07,berg07}. In the present case, the crystal structure is orthorhombic and the stripes are rotated from the Cu-O bond direction, so that a finite interlayer Josephson coupling is allowed (and observed \cite{dord03}).  Hence, the presence of a PDW superconducting glass state in our LSCO sample is quite plausible.

\subsection{Inequivalent oxygen sites and stripe pinning}

The onset of charge and spin stripe order in LBCO is limited by a transition to the LTT (or LTLO) structure that results in inequivalent oxygen sites within the CuO$_2$ planes \cite{huck11}, as illustrated in Fig.~\ref{fg:io}.    Hence, it has been surprising that in LSCO, with a presumed average LTO structure and corresponding equivalency of all in-plane O sites \cite{rada94}, the NMR evidence for slow spin and charge fluctuations begins at temperatures even higher than in LBCO \cite{hunt99,baek12b,baek14}.  While our evidence for a lowered LTLO symmetry resolves this puzzle, it is of interest to consider relevant experimental results from the past.

In an early study of LSCO with $x=0.14$, Nohara {\it et al.} \cite{noha93} found a softening of the transverse elastic constant $(c_{11}-c_{12})/2$ below 50 K.  They noted that the associated strain on the lattice would lead to inequivalent oxygen sites, as in the LTLO structure.  Later tests of structure tended to focus on $x=0.12$.  Transmission electron microscopy (TEM) studies demonstrated the presence of (110)-type superlattice peaks, consistent with either the LTLO phase or LTT phase \cite{koya95,hori97,hori00}.  Imaging these peaks revealed that the lower-symmetry phase is present at the domain boundaries between twins of the LTO phase.  A synchrotron x-ray powder diffraction study by Moodenbaugh {\it et al.} \cite{mood98} found evidence for a small fraction ($\sim10$--20\%) of LTT phase within the LTO phase of LSCO x=0.12 for $T\lesssim100$~K \cite{mood98}.  For this same composition, an NMR study reported a change in symmetry below 50~K, consistent with LTT-like tilts \cite{goto08}.  
 
In the present study, as the neutron diffraction measurement averages over the sample volume, we have to allow the possibility that the weak $(3\bar{3}0)$ superlattice peak could be coming from a small fraction of the sample, such as twin boundaries.  At the same time, the strength of the octahedral-tilt phonons indicates a response from the entire volume.  That a small fraction of these fluctuations might condense throughout the volume should not be surprising.  In fact, there might be a small tilt ordering in the bulk and a much larger ordering in the twin domains.  A local-probe measurement, such as TEM, will be necessary to provide a definitive answer.
 
Previous studies of soft phonons in LSCO \cite{lee96,waki04b} have focused on the octahedral tilts about plaquette diagonals.  In the high-temperature tetragonal (HTT) phase, there are two such modes that are orthogonal and equivalent.  One of these effectively goes soft at the transition to the LTO phase \cite{axe89}, resulting in superlattice peaks of the type (032).  The possible softening of the other mode can, in principle, be studied at positions such as (302); however, the inevitable presence of twin domains means that the (302) of one domain is nearly superimposed on the (032) of its twin, complicating the measurements.   On the other hand, a position such as (330) is a forbidden superlattice reflection in the LTO phase,  but is allowed in the LTT and LTLO phases \cite{huck11}.  As already mentioned, soft phonons at this position directly correspond to displacements that cause in-plane oxygens to be inequivalent.   In a recent study of LBCO with $x=0.125$ \cite{bozi15}, soft phonons were observed at (330) with an integrated intensity that was independent of temperature in the higher temperature phases (HTT and LTO); on entering the LTT phase at 54 K, the intensity of the soft phonons was transferred to the superlattice peak at zero energy.

For LBCO, theory finds, correctly, that the ground state has static LTT tilts, but that the LTO structure is favored at higher temperatures due to the free-energy gain from the higher entropy of that state \cite{pick91,cai94}.  The LTT tilts still cost little energy, but contribute entropy as they are unable to order in the LTO phase.  For our LSCO $x=0.07$ sample, we appear to have an intermediate situation: a small fraction of the LTT-like tilt character condenses, but most of it is prevented from doing so by the orthorhombic symmetry, which likely becomes LTLO instead of LTO.  The LTT-like tilt fluctuations have a quite noticeable impact on the thermal conductivity.  In the case of LBCO, complete or nearly complete ordering of the tilts results in a substantial increase in the thermal conductivity below the transition temperature \cite{wen12a}.  In contrast, thermal conductivity measurements on LSCO with $x\geq0.1$ show no significant recovery at low temperature \cite{naka91,sun03}, consistent with the continued presence of dissipative fluctuations.

The impact of the LTT-like phonons should have only an indirect impact on the spin fluctuations.  Condensing some of these phonons to induce the LTLO phase allows charge stripes to be pinned, which in turn allows the spin stripes to become quasistatic.  In contrast, a different phonon mode has recently been shown to couple directly to the spin fluctuations \cite{wagm15,wagm15b}; it involves in-plane displacements of O atoms in directions transverse to the Cu-O bonds \cite{wagm15}.

It is interesting to note that Baledent {\it et al.} \cite{bale10} studied a weak (110) peak, equivalent to our (330), that appeared below 120~K in an LSCO crystal with $x=0.085$ and $T_c=22$~K.  They observed this feature with polarized beam in the spin-flip channel, and hence interpreted it as a magnetic diffraction peak.   Given the present results regarding structural correlations, it may be worthwhile to test the degree of magnetic character at (110).

\subsection{Rotation of incommensurate peaks}

The rotation of the IC magnetic peaks away from the Cu-O bond direction is consistent with previous measurements on LSCO $x=0.12$ \cite{kimu00,tham14} and La$_2$CuO$_{4+\delta}$ \cite{lee99}; it is also consistent with the orthorhombic symmetry, as discussed in \cite{robe06}.  The magnitude of the rotation found here for $x=0.07$, $\sim 14^\circ$, is significantly larger than that reported for $x=0.12$ ($\sim3^\circ$).  For $x\sim0.055$, the rotation reaches $45^\circ$ \cite{fuji02c}, with the spin modulation occurring uniquely along the $b$ axis in the spin-glass regime \cite{waki00}.

The rotation of the modulation direction at $x\sim 0.055$ corresponds with the superconductor to insulator transition.  A careful study of resistivity in an LSCO thin film with carrier concentration varied electrostatically provides evidence that the transition involves a localization of pairs \cite{boll11}.  A torque magnetometry study found evidence that superconducting fluctuations survive on the insulating side of the transition \cite{li07b}.  On the larger $x$ side, measurements of hysteresis in magnetoresistance indicate charge glass character of the insulator state, with residual effects observable in a superconducting $x=0.06$ thin film \cite{shi13}.   Even in zero field, there is upward curvature of $\rho_{ab}(T)$ at $T\lesssim100$~K for $x\lesssim0.13$ that grows as $x$ decreases toward 0.06 \cite{komi05}.

Hence, it appears that the rotation of the IC peaks in our $x=0.07$ sample may involve a mixing of the characters, superconducting and insulating, respectively associated with the bond-parallel and plaquette-diagonal modulations.  The insulating character need only involve localization, rather than the breaking, of pairs.    As doping increases, the proportion of diagonal/insulating character decreases, as indicated by the reduced rotation angle found for $x=0.12$ \cite{kimu00}.  

\subsection{Connection with nodal gap?}

The presence of a spatially-modulated superconducting state should have an impact on the observed superconducting gap.  When the superconductivity is uniform, the gap is observed to have $d_{x^2-y^2}$ symmetry, with gapless nodes along the $(1,1)$ and $(1,-1)$ directions \cite{tsue00}.   Interestingly, anomalous behavior has been detected in LSCO  with $x \lesssim0.1$.   In a recent angle-resolved photoemission spectroscopic (ARPES) study of LSCO $x=0.08$  \cite{razz13}, the appearance of a gap at the nodal points was inferred, starting at temperatures as high as $\sim 80$~K and growing on cooling.  Given that the resistivity on such a sample does not show either strong insulating or superconducting behavior at such high temperatures, the nature of the gap is somewhat ambiguous; nevertheless, there is a clear change in the spectral function at the nodal wave vector as a function of temperature.  

To put this in perspective, we can compare with the case of LBCO with $x=1/8$, where the presence of PDW order has been inferred \cite{li07,berg07}.  Theoretically, the PDW superconductor is predicted to have a large gap in the antinodal regions, but to have no gap along an arc of states centered on the nodal direction \cite{berg09a,baru08}.  Experimentally, a $d$-wave like gap is seen in ARPES measurements about the nodal regime \cite{vall06,he09}; the deviation from the PDW prediction could be due to the presence of spin stripe order, which was not included in the theory for the gap \cite{frad15}.

Further evidence that the gap in underdoped LSCO is different from that in LBCO has been provided by studies of thermal conductivity.  At low temperature, the thermal conductivity, $\kappa$, is dominated by the contribution from electrons, and in the presence of a $d$-wave gap, only quasiparticles near the nodes can contribute.  It has been shown that the ratio of the thermal conductivity to temperature, $\kappa/T$, in the limit $T\rightarrow 0$, has a universal form that depends only on the Fermi velocity and the gap velocity near the node \cite{durs00}.  Low-temperature measurements of $\kappa$ have provided measures of the superconducting gap in a variety of cuprates that are consistent with results from ARPES \cite{suth03}; however, anomalous behavior is observed in LSCO with $x\lesssim0.14$ \cite{suth03,hawt03,sun03c}.  Effectively, the thermal conductivity is smaller than one would expect from well-defined quasiparticles, and the magnitude decreases in a magnetic field, contrary to conventional behavior.  

One way to explain the results for underdoped LSCO is to introduce a small gap at the nodal point \cite{gusy04}; however, to explain the response to a magnetic field, one must then assume that the gap grows with field.  Alternatively, it has been proposed that the field induces local antiferromagnetic order within the cores of the field-induced vortices, thus suppressing the quasiparticles essential for heat transport at low temperature \cite{taki04}.  The case of LBCO with $x=1/8$ is an example where local antiferromagnetic order, in the form of spin stripes, does {\it not} result in a nodal gap \cite{huck11,he09}.  A difference in the case of LSCO is the rotation of the stripe orientation.  The mixing in of diagonal stripe character might be responsible for the apparent nodal gap \cite{zhou15}. 

\section{Summary}
\label{sc:summary}

We have presented a neutron scattering study of LSCO $x=0.07$, an underdoped cuprate with bulk superconductivity and $T_c=20$~K.  We have observed incommensurate spin fluctuations with a gapless spectrum that coexists at low temperature with the superconductivity.  These results are consistent with a periodic modulation of the superconducting state intertwined with the spin order \cite{frad15}, and they demonstrate that the related behavior previously observed in LBCO \cite{li07,tran08,xu14} is not unique.  We have also discovered evidence of a lowering of symmetry below that of the presumed LTO phase, likely to the LTLO; in addition, considerable entropy remains in LTT-like phonons at low temeperature.  These observations provide a connection to the stripe pinning found in the LTT phase of LBCO \cite{huck11}.   The weakly-ordered spin stripes are rotated away from the Cu-O bond directions, consistent with previous studies of LSCO with $x=0.12$ \cite{kimu00,tham14}, and suggesting pinning of charge stripes in the proposed LTLO phase.  We have pointed out that the mixing in of some diagonal stripe character may help to explain a recent ARPES observation of a nodal gap that appears even above $T_c$ \cite{razz13}.

\acknowledgments

For the work at HYSPEC, we are grateful to M. K. Graves-Brook for valuable assistance.  Helpful comments from S. A. Kivelson, R. M. Konik, T. Senthil, and A. M. Tsvelik are gratefully acknowledged.  H.J. was supported by the Danish Research Council FNU through DanScatt.  Work at Brookhaven was supported by the Office of Basic Energy Sciences (BES), Division of Materials Sciences and Engineering, U.S. Department of Energy (DOE), through Contract No.\ DE-SC00112704.   This work utilized facilities at the NCNR supported in part by the National Science Foundation under Agreement No.~DMR-0944772.  The experiments at ORNL's SNS and HFIR were sponsored by the Scientific User Facilities Division, BES, U.S. DOE.  


%

\end{document}